# Rotational and translational drags of a Janus particle close to a wall and a lipid membrane

Vaibhav Sharma,[1] Florent Fessler,[1] Fabrice Thalmann,[1] Carlos M Marques,[2] Antonio Stocco[1,*]

[1]Institut Charles Sadron, CNRS UPR22—University of Strasbourg, 23 rue du Loess, Strasbourg, 67034, France
[2]ENS Lyon, CNRS, Université Lyon 1, Laboratoire de Chimie UMR 5182, F-69342, Lyon, France
*stocco@unistra.fr

## Abstract

*Hypothesis:* Measuring rotational and translational Brownian motion of single spherical particles reveals dissipations due to the interaction between the particle and the environment.
*Experiments*: In this article, we show experiments where the in-plane translational and two rotational drag coefficients of a single spherical Brownian particle can be measured. These particle drags are functions of the particle size and the particle-wall distance, and of the viscous dissipations at play. We measure drag coefficients for Janus particles close to a solid wall and close to a lipid bilayer membrane.
*Findings*: For a particle close to wall, we show that according to hydrodynamic models, particle-wall distance and particle size can be determined. For a particle partially wrapped by lipid membranes, in absence of strong binding interactions, translational and rotational drags are significantly larger than the ones of non-wrapped particles. Beside the effect of the membrane viscosity, we show that dissipations in the deformed membrane cap region strongly contribute to the drag coefficients.

## 1 Introduction

Increase of the particle drag in fluid confinement occurs for many systems ranging from surfactant molecules inside porous materials to colloidal particles close to a solid-liquid interface or a membrane. The presence of an interface impacts the hydrodynamic flow around a translating or rotating particle, and the interaction between a particle and a surface implies boundary conditions for the fluid flow at the interface and emerging dissipations contributing to the drag [1].

Hydrodynamic calculations by Faxén addressed the translational and rotational drags of a spherical particle as a function of the particle distance to a solid wall [2]. Goldman and Brenner [3] extended the description for the particle drag and many other hydrodynamic models can be found in the literature for the drag of a particle close to liquid-liquid and liquid-gas interfaces [4,5][6][7]. Of particular interest in membrane biophysics, the Saffman-Delbrück model describes the dissipation of a disk particle embedded in a flat viscous membrane between two unbounded fluids [8]. Studies related to this model also addressed the rotational drag of a disk in a membrane [9] and the effect of a solid wall close to the membrane [10]. Membrane shear and dilatational viscosities together with the related adimensional Boussinesq numbers are used to describe the competition between the viscous surface and bulk dissipations. Surface flow compressibility conditions must be also carefully taken into account when dealing with lipid membranes or surfactant monolayers.

To verify theoretical models and to evaluate the dissipations at play, experimental investigations using colloidal particles close to solid interfaces or membranes have been reported [11][12]. Experiments are very challenging in particular when the gap distances between the particle and the wall decrease to the nanometric scale [13]. Experimental reports on particles interacting with lipid membranes reveal that, beyond viscous dissipations, an hydrodynamic coupling exists between the particle and the fluctuating elastic membrane [14][15][16][17][18,19].

We have recently reported on the motion of active and passive Janus colloids close to giant unilamellar vesicle (GUV) membranes and we have shown two preliminary experiments on the out-of-plane rotational dynamics of Janus particles partially wrapped by lipid membranes [20][21]. In reference [21] we focussed our attention on the preferential orientations of Janus particles with respect to the membrane surface, which were discussed in terms of interfacial energies. Here, we report experiments on the translational, in-plane and out-of-plane rotational drags, which are simultaneously measured for single Janus colloids. In the first part of the Results section, we describe experiments for single particles close to a solid wall, and show a good agreement with the Faxén predictions. Then, we present experimental results for Janus particles that are also partially wrapped by a lipid membrane and discuss the origin of the dissipations contributing to each particle drag coefficient.

## 2 Materials and Methods

### 2.1 Janus Colloids

We used bare fluorescent Melamine Formaldehyde Resin (MF) microspheres of 1.245 µm nominal radius and nominal standard deviation of 0.05 µm (microParticle GmbH, Berlin, Germany). To fabricate Janus colloids we followed the method proposed by Love *et al.*[22]. First, a monolayer of MF beads was deposited on a thoroughly cleaned silica wafer by drop casting (particle concentration of 0.2% by volume). After completely drying, a thin layer of platinum was deposited on the colloidal monolayer using metal sputtering (Auto 306 Evaporator, BOC Edwards, West Sussex, UK). Due to the spherical geometry, the sputtering process yields particles that are half coated with a thin platinum layer with a thickness of 6 ± 1 nm, measured by light reflectivity on the corresponding planar surfaces. For these Janus particles, we performed scanning electron microscopy and measured an average radius $R_P$ = 1.29 µm and a standard deviation of 0.06 µm. The Janus colloids are released from the wafer by simple agitation using a pipet tip[20][21].

Zeta potential $Z$ of the MF and MF-Pt Janus particles was measured using a Malvern Zetasizer Nano ZS: $Z_{MF}$=+25±6 mV for MF particles and $Z_{MF-Pt} = -47±10$ mV for MF-Pt Janus particles.

### 2.3 GUV formation

GUV's are prepared using PVA (polyvinyl alcohol) gel assisted formation method, which enables high-yield vesicle growth [23]. In this method, we first start with preparing PVA gel by dissolving dry PVA in PBS (phosphate-buffered saline, 10gL⁻¹) solution at 5% concentration. The PVA gel is then spread inside the chambers of a homemade PTFE (polytetrapolyethylene) plate and oven dried at 80°C for 30 min. Following the drying process, a 99:1 (molar) mixture of POPC-NBD (1-palmitoyl-2-oleoylphosphatidylcholine fluorescently labelled with nitrobenzoxadiazole) in chloroform (1 gL⁻¹) is spread on the dried PVA gel and then vacuum dried for 15 min in a desiccator. The lipids are next hydrated with 200 mL of 0.15 M sucrose and let to grow for at least 2-3 hours. After growing, the vesicle suspension is collected and sedimented in 1 mL of 0.15 M glucose solution. Because of the small density mismatch between the solutions inside the vesicle and outside in the aqueous media, the vesicles settle to the bottom of the collection tube. The average GUV radius measured after gel assisted formation method was found to be $R_{GUV}$ = 11 ± 6 µm.

### 2.4 Microscopy and tracking

Janus colloids and GUVs in aqueous solution were observed using fluorescent microscopy. The sample cell is filled with 10 µL of Janus particle and GUV solution, along with 120 µL of glucose. The microscopy setup consisted of Nikon (Tokyo, Japan) Eclipse TE2000 microscope (x60 objective) equipped with a CMOS camera (Orca Flash 4.0, Hamamatsu, Japan). Videos were recorded at different frame rates ranging from 100 to 1000 frames per second (fps). Using the open-source software Blender v2.8 (Blender Foundation, Amsterdam, The Netherlands), tracking the center of mass of the particle was achieved. To track the orientation of the colloids a thresholding technique using ImageJ (NIH, Bethesda, MD, USA) was used [21]. The analysis of partially coated MF-Pt colloids yields a brighter area $A_{MF}$, which corresponds to the MF part of the Janus colloid observed in reflection mode of florescence microscopy. The area $A_{MF}$ provides the orientation angle $\beta$: $A_{MF} = \pi/2(1 - \cos(\pi - \beta))R_P^2$ (See Figure 1A and 1B right) [21]. Moreover by fitting the perimeter of this area with an ellipse, the in-plane orientation $\varphi$ can be evaluated (See Figure 1B left).

## 3 Results and Discussion

### 3.1 Translational and Rotational drags of single particles close to a wall

We started investigating the Brownian motion of single MF-Pt Janus colloids in thermal equilibrium close to the solid-liquid interface, where particles sediment because of gravity but do not adsorb on the substrate (given the overall negative potential of the particle and the wall, see 2.1 section). Very low particle concentrations allow to study the motion of single isolated particles. For each particle, we were able to measure the center of mass position and the orientation of the Pt region of the particle. By performing high speed imaging we focused our attention on relatively small variation of $\beta$ and $\varphi$ - the particle orientation angles - to measure dynamics related to Brownian diffusions $D_i = k_BT/\zeta_i$, where $k_BT$ is the thermal agitation energy and $\zeta_i$ is the drag coefficient [24].

Fig. 1A and 1B sketches the geometry of our experimental system composed of a Janus particle close to a wall together with the $xyz$ laboratory axis and the $XYZ$ particle axis. In Fig. 1C, typical images recorded in fluorescence microscopy used to measure the in-plane and out-of-plane particle orientations, $\varphi$ and $\beta$, are shown.

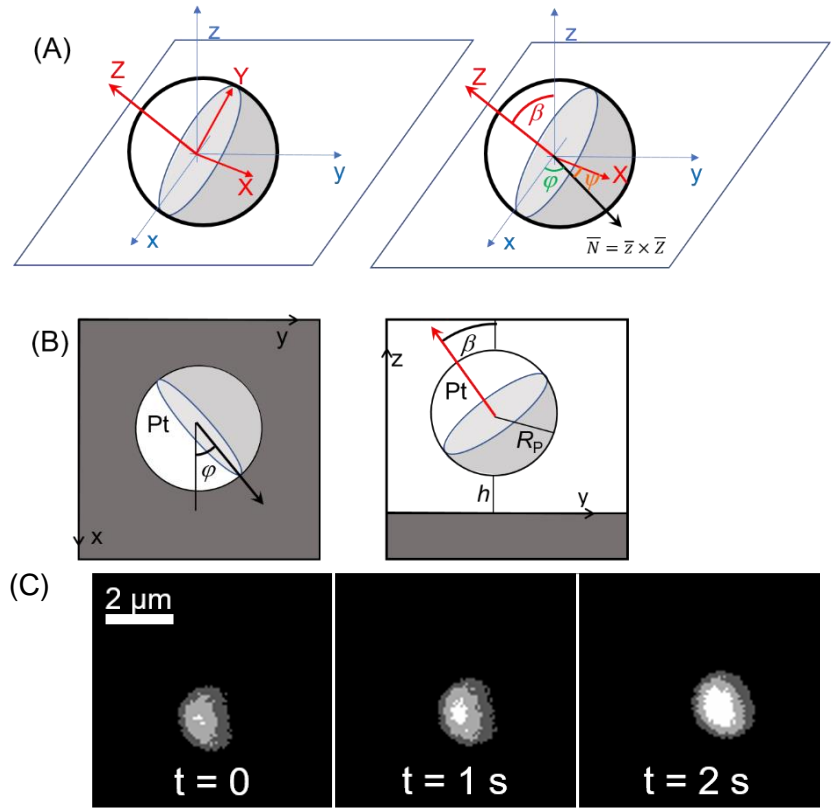

Fig. 1. (A) 3D Sketches of a Janus particle in the $xyz$ lab frame and $XYZ$ particle frame and definition of the Euler angles $\varphi\beta\psi$. (B) 2D sketches of a top view ($xy$-plane) and side view ($xz$-plane) of a colloid close to a solid wall, and definition of the particle orientation angle $\varphi$ and $\beta$. (C) Florescence images at given times of a $R_P$ = 1.29 ± 0.16 µm Janus particle ($xy$-plane).

A translational trajectory and variations of the in-plane and out-of-plane orientation angles as a function of time are shown in Fig. 2A.

Time averaged mean squared translational displacement *MSD* and mean squared angular displacements *MSAD*s ($<\Delta\varphi^2>$ and $<\Delta\beta^2>$) of the corresponding trajectories are plotted in Fig. 2B as a function of the lag time $\Delta t$ [25].

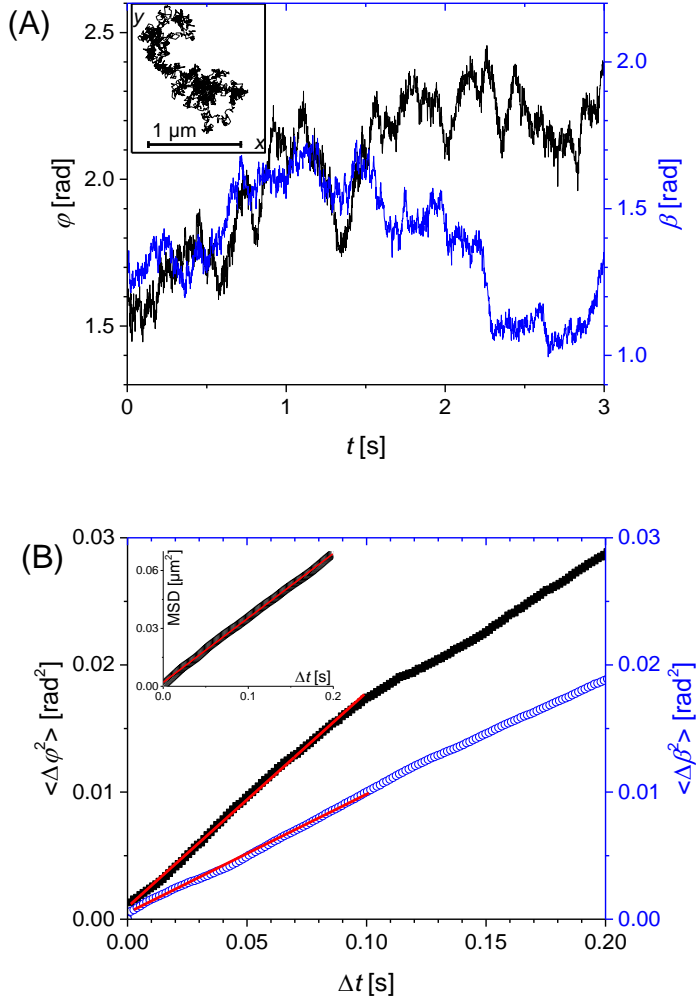

Fig. 2. (A) In plane (left axis) and out of plane (right axis) particle orientations as a function of time. Inset shows the center of mass trajectory. (B) Mean squared angular displacement corresponding to the orientation angles shown in (A) as a function of the lag time. Inset shows the mean squared displacement for the center of mass trajectory. Lines correspond to the fits of equation 1 to the data in the short lag time limits.

In the short lag time limit, mean squared values allow measuring the translational drag parallel to the wall $\zeta_{tr,\parallel}$ and the in-plane $\zeta_{ro,\parallel}$ and out of plane $\zeta_{ro,\perp}$ rotational drags close to a wall ($D_i = k_B T/\zeta_i$). For the translational motion, the vertical motion (in the *z*-axis) is strongly confined by the potential energy of the particle, which can be modelled as the sum of gravity and electrical double layer repulsion terms [26]. In these conditions, only few tens of nm can be explored by the Brownian particle in the *z*-axis. In the horizontal plane, instead, the particle experiences free Brownian motion. Hence, only the motion parallel to the wall in the *xy* plane is considered here:

$$MSD = <\Delta x_P^2> + <\Delta y_P^2> = <\Delta L^2>_0 + 4(k_B T/\zeta_{tr,\parallel})\Delta t \qquad (1A)$$

For the rotational dynamics, three particle rotations about three independent axis related to the Euler angles in our geometry must be taken into account (see Fig. 1A) [27]. Considering the particle orientation $\varphi$ and $\beta$ measured in our experiments, mean squared angular displacements read (see Supplementary Material, S1):

$$< \Delta\beta^2 > = < \Delta\beta^2 >_0 + 2(k_B T/\zeta_{\mathrm{ro},\perp})\Delta t \qquad (1B)$$

$$< \Delta\varphi^2 > = < \Delta\varphi^2 >_0 + 2k_B T(1/\zeta_{\mathrm{ro},\parallel} + \cot^2 < \beta >/\zeta_{\mathrm{ro},\perp})\Delta t \qquad (1C)$$

where an offset ($< \Delta L^2 >_0$ or $< \Delta\varphi^2 >_0$ or $< \Delta\beta^2 >_0$) due to experimental errors (e.g. noises in image acquisition, image analysis or mechanical vibrations) is always taken into account [28]. In equation 1C, note that for $< \beta > \neq \pi/2$ both rotational friction coefficients should be taken into account as explained in our theoretical analysis (see Supplementary Material, S1). The in-plane rotational drags are very close to bulk values calculated with a particle radius $R_P$= 1.29 ± 0.16 µm (see later in the text, Figure 6), which agree with the hydrodynamic predictions [13].

$$\zeta_{\mathrm{ro},\parallel} \cong \zeta_{\mathrm{ro,b}} = 8\pi R_P^3 \eta \qquad (2)$$

On the contrary, translational and the out-of-plane rotational drags are significantly smaller than the calculated bulk values [13].
Hydrodynamic calculations by Goldman *et al.* provide translational $\zeta_{\mathrm{tr},\parallel}$.and rotational $\zeta_{\mathrm{ro},\perp}$ particle drags as a function of the particle-wall distance [13][3]. Here, we compare our experimental data with these predictions in Fig. 3 where $\zeta_{\mathrm{ro},\perp}$ is plotted as a function of $\zeta_{\mathrm{tr},\parallel}$.
In Fig. 3 our data agree with the hydrodynamic predictions for $R_P$ = 1.29 ± 0.16 µm and gap distance $h$ in between $0.1R_P$ and $1R_P$. This large distribution of $h$ could be discussed by considering the Janus geometry of the particles and the heterogeneous surface potential due to the native positive value of MF ($Z_{\mathrm{MF}}$ = +25 mV) and the negative value of the Pt coating ($Z_{\mathrm{Pt}}$ = −80 mV [29]) leading to a negative potential of the Janus colloids: $Z_{\mathrm{MF-Pt}}$ = −47 ± 10 mV.
Note that the range of the particle radius used in this comparison (± 0.16 µm, Fig. 3) is larger than the one by SEM measurements (± 0.06 µm). The difference between the measurements of the hydrodynamic radius by particle tracking and particle radius by electron microscopy can be related to the Janus particle geometry, heterogeneous particle surface properties and distribution of electrolytes close to the particle surface [30].
It is important to remark that by measuring both $\zeta_{\mathrm{tr},\parallel}$ and $\zeta_{\mathrm{ro},\perp}$ for the same particle we can in principle evaluate both the size and the particle distance to the wall, if the particle drags are only due to hydrodynamics [1]. Significant errors can be made when the particle distance to the wall is evaluated by measuring only $\zeta_{\mathrm{tr},\parallel}$ since the size of the particle must be assumed and even a small size distribution affects strongly the evaluation of $h$, see Fig. 3.

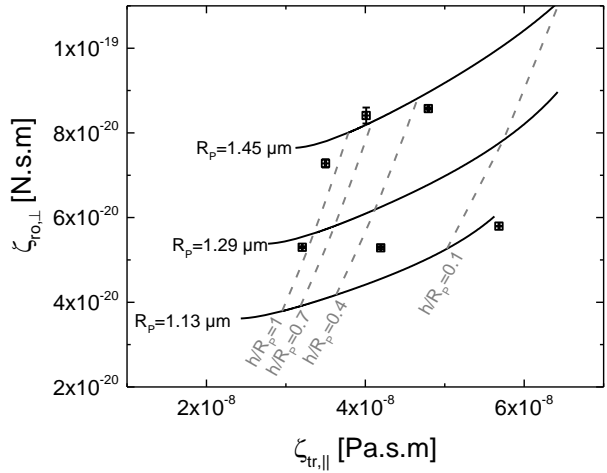

Fig. 3. Out of plane rotational drag as a function of the parallel translational drag: experimental points (square) are compared to the hydrodynamic predictions (lines) for different radii and relative distances to the wall $h/R_P$.

### 3.2 Translational and Rotational drags of partially wrapped particles by lipid membranes

As reported recently, we were able to drive the wrapping of giant vesicle lipid membranes on Janus particles by applying external forces of the order of 10 pN (see Supplementary Material, S2) [21][14]. In fluorescence microscopy, many MF-Pt Janus colloids show preferential orientations with the fluorescent MF region facing the bottom of the cell as described in reference [21], but some show averaged orientations comparable to the ones of non-wrapped particles (see Supplementary Material, S2). We sketched this geometry in Fig. 4A, where $d$ is the gap distance between the particle and the membrane and $d_w$ is the gap distance between the membrane and the glass substrate [31–34].

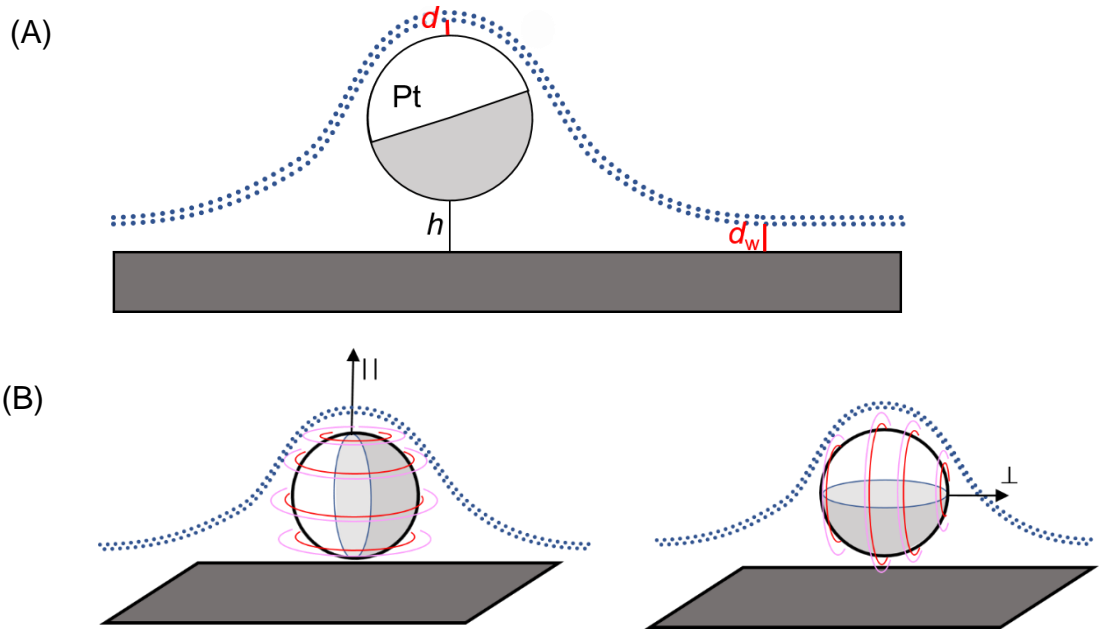

Fig. 4. Sketches of a Janus colloid partially wrapped by a lipid membrane. (A) The membrane-particle gap distance is $d$, membrane-substrate gap distance is $d_w$. (B) Sketches of the particle rotational dynamics related to the parallel (∥) and perpendicular (⊥) rotational drag coefficients.

As before, we were able to monitor the particle position and orientations and calculate mean squared angular displacements and translational mean squared displacement as a function of the lag time, which are shown in Fig. 5.

To assess the precision of our analysis, we also plot data for a Janus particles stuck on the solid substrate and show that *MSAD* and *MSD* do not significantly vary with the lag time (x points).

For partially wrapped Janus colloids we performed the same analysis as before to evaluate the translational drag and the in-plane rotational and out-of-plane rotational drags. The effect of confinement was taken into account when $< \Delta \beta^2 >$ shows a non linear behaviour at short lag times (see filled circles in Fig. 5B). In this case, the mean squared angular displacement data were fitted by [29]:

$$< \Delta \beta^2 > = < \Delta \beta^2 >_0 + \frac{k_B T / \zeta_{\text{ro},\perp}}{\Gamma} [1 - \exp(-2\Gamma \Delta t)] \tag{3}$$

Where $\Gamma$ acts as an effective elastic constant, which in this case, tends to restore the particle orientation $\beta$ to a value corresponding to the minimum interaction energy between the membrane and MF-Pt particle [21].

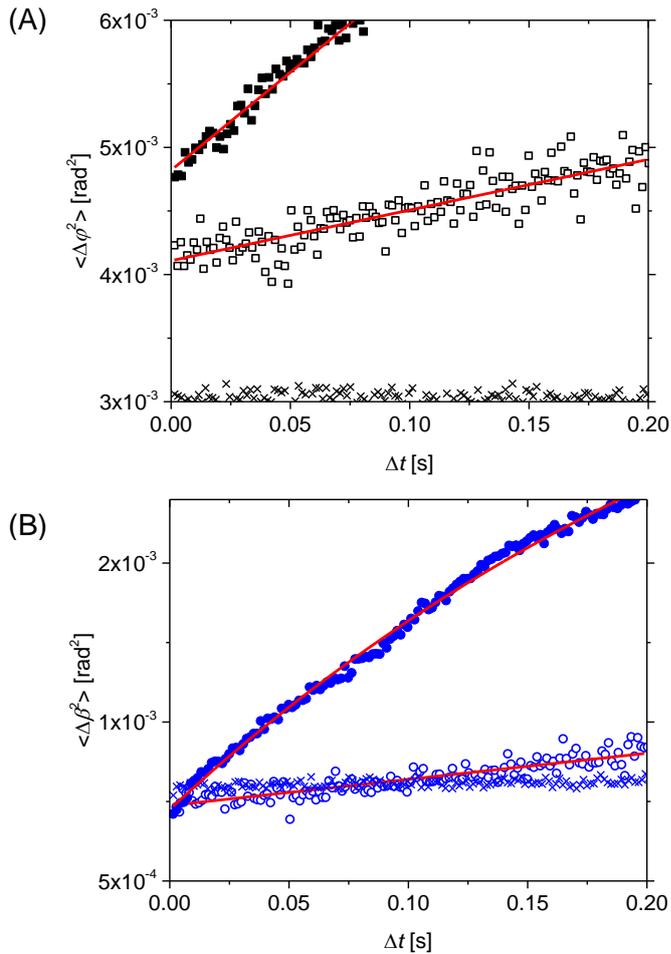

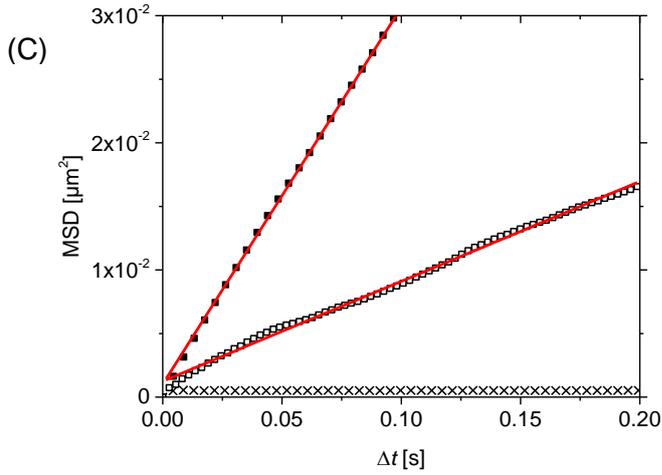

(C)

Fig.5. For two particles partially wrapped by a membrane (empty and filled squares) and for a particle stuck on the solid substrate (x): (A) Mean squared in-plane angular displacements, (B) mean squared out-of-plane angular displacements, and (C) mean squared translational displacement MSD as a function of the lag time (not all points are shown for clarity). Lines correspond to the fits of equation 1 and 3 to the data in the short lag time limits.

In Fig. 6 we plot $\zeta_{ro,\perp}$ as a function of $\zeta_{ro,\parallel}$ for "free" (non-wrapped) Janus particles close to a wall and for Janus particles partially wrapped by a giant vesicle membrane. For "free" particles $\zeta_{ro,\perp}$ is slightly higher than $\zeta_{ro,\parallel}$ but their values are comparable to the bulk values (blue cross points for $R_P$ = 1.13, 1.29 and 1.45 µm).

For "wrapped" particles $\zeta_{ro,\perp}$ is significantly higher than $\zeta_{ro,\parallel}$ and the two rotational drags show a clear correlation that can be approximately described by $\zeta_{ro,\perp} \approx g\, \zeta_{ro,\parallel}$ , with $1 < g < 3$. Note that for wrapped particles, rotational drags can be 1 or 2 orders of magnitude higher than for non-wrapped particles.

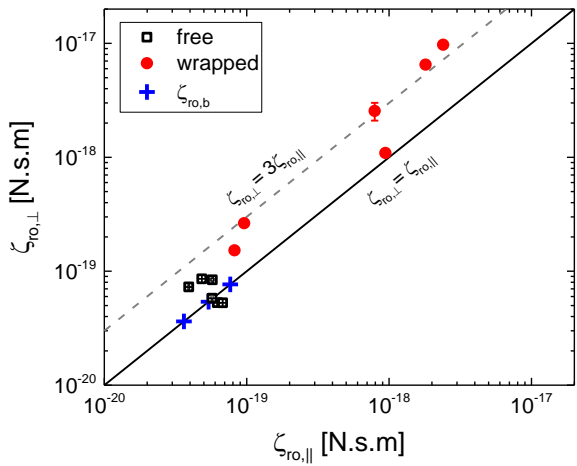

Fig.6. Out of plane rotational drag as a function of the in plane rotational drag: experimental points for "free" particles (square) and for particles partially "wrapped" by a GUV membrane (circles). Bulk rotational drag values for $R_P$ = 1.13, 1.29 and 1.45 µm are also shown (+) for comparison.

Now, we discuss the origin of the dissipations contributing to each particle drag coefficient. Rotational drag $\zeta_{\mathrm{ro},\parallel}$ experienced by a Janus particle partially engulfed by a membrane and rotating about an axis parallel to the surface normal (as sketched in Fig. 4B) can be described as the combination of several contributions.

Contribution of membrane viscosity can be compared to the one of a rotating disk embedded in a flat membrane and close to a solid wall [10]:

$$\zeta_{\mathrm{ro,D}} \cong 4\pi R_{\mathrm{P}}^2 \eta_{\mathrm{m}} \qquad\qquad\qquad (4)$$

The contribution of the fluid viscosity for the unwrapped region of the particle can be approximated as a fraction of the bulk contribution $\zeta_{\mathrm{ro,b}}$ for a spherical particle considering the degree of wrapping. The remaining contribution to $\zeta_{\mathrm{ro},\parallel}$ concerns the dissipation occurring in the membrane cap region of the particle. This region includes both the area wrapped by the membrane, which is not in direct contact with the particle but separated by a water gap, and the deformed membrane region close to the particle [35][36]. Hence, the membrane cap contribution $\zeta_{\mathrm{ro,cap}}$ takes into account all the dissipations beyond the assumption of a flat membrane and a particle inclusion.

Total drag $\zeta_{\mathrm{ro},\parallel}$ may be written as a function of these contributions:

$$\zeta_{\mathrm{ro},\parallel} \cong \mathcal{F}(\zeta_{\mathrm{ro,D}}, \zeta_{\mathrm{ro,b}}, \zeta_{\mathrm{ro,cap},\parallel}). \qquad\qquad (5)$$

Similarly, for the out-of-plane rotational drag ($\perp$ as sketched in Fig. 4B) we can write:

$$\zeta_{\mathrm{ro},\perp} \cong \mathcal{F}(\zeta_{\mathrm{ro,b}}, \zeta_{\mathrm{ro,cap},\perp}) . \qquad\qquad\qquad (6)$$

Given the correlation $\zeta_{\mathrm{ro},\perp} \approx g\, \zeta_{\mathrm{ro},\parallel}$ shown in Fig. 6 and considering typical literature values of $\eta_{\mathrm{m}} = 2\ 10^{-9}$ Pa.s.m [14], we can deduce that for our particle size ($R_{\mathrm{P}} \approx 10^{-6}$ m) the contribution $\zeta_{\mathrm{ro,D}}$ is comparable to the bulk contribution ($4\pi R_{\mathrm{P}}^2 \eta_{\mathrm{m}}$ *vs* $8\pi R_{\mathrm{P}}^3 \eta$). These contributions are small compared to the dissipations measured for particles wrapped by the membranes, see Fig. 6. Hence, both $\zeta_{\mathrm{ro},\perp}$ and $\zeta_{\mathrm{ro},\parallel}$ mainly reflect the contributions arising from the dissipations in the deformed membrane region close to the particle, which are not described in the existing models assuming a flat membrane and a particle inclusion.

$\zeta_{\mathrm{ro},\perp} > \zeta_{\mathrm{ro},\parallel}$ can be explained in terms of a wall effect as it is true for "free" (non-wrapped) particles close to a wall, which also points to a very small particle-wall distance $h$ in our experiments.

To the best of our knowledge, no models exist in the literature to describe $\zeta_{\mathrm{ro,cap}}$ in our system. In the membrane cap region, shear stresses are acting on the water gap, the membrane and also in the bulk water on the other side of the membrane. For the particle rotational dynamics, it is important to realize that the relevant lipid bilayer membrane shear viscosity may be different far from the cap and in the cap region. For the parallel rotational drag, far from the cap the gradient of the velocity is along a direction that lays in the membrane plane; whereas in the cap region at the particle equator the gradient of the velocity is normal to the membrane plane (Fig. 4B). In the latter case an inter-leaflet viscosity could also play a role, which describes an additional dissipation due to the sliding between the two monolayers of the lipid bilayers [1].

For the sake of comparison, we could refer to some limiting scenarios to describe our data. First, we consider the case of an immobile membrane and refer to the hydrodynamic result for a spherical Couette geometry. Dissipation is due to the slow motion of the fluid contained (in the small gap) between two concentric hard spheres by assuming that the inner sphere rotates while the outer sphere stays at rest. Hence, we can assume that the radius of the inner sphere is $R_{\mathrm{P}}$ and the radius of the outer sphere is $R_{\mathrm{P}} + d$ (Fig. 4B). Note that the lipid membrane is assumed to act as a solid-liquid interface, and no dissipation due to the membrane viscosity can be described within this assumption. In this case, the rotational drag felt by the inner sphere can be written as [37]:

$$\zeta_{\text{ro,S}} = \frac{\zeta_{\text{ro,b}}}{[1-(\frac{R_P}{R_P+d})^3]} = \frac{8\pi\eta R_P^3}{[1-(\frac{R_P}{R_P+d})^3]} \tag{7}$$

Hence, plotting $\zeta_{\text{ro,S}}$ as a function of $d$ in Fig. 7A we note that a variation of the gap distance $d$ from 10 nm to 600 nm results into frictions that agree with the $\zeta_{\text{ro,||}}$ experimental results. The distance $d$ between the spherical microparticle and the membrane can be compared to the equilibrium distances $d_w$ measured between planar solid substrates and vesicle membranes in the absence of strong binding interactions, which usually lay in the range between 20 and 60 nm [31–34].

Alternatively, for a flat viscous membrane we can discuss the limiting case $\zeta_{\text{ro,||}} \cong \zeta_{\text{ro,D}} + \zeta_{\text{ro,b}}$ where no internal dissipation due to the flow in the gap between the particle and the membrane occurs. Fig. 7B shows that our experiments agree to the calculations for a sphere with an effective radius $R_{\text{eff}}$ ranging from 1.29 µm (= $R_P$) to 4.3 µm. The latter large effective radius may describe the contribution of the deformed membrane area close to the particles. A typical length scale that defines this area is the bendocapillary length $\lambda = \sqrt{\kappa_B/\sigma}$ of the membrane. In our experiments, the bending rigidity is $\kappa_B \cong 10^{-19}$ J; and the membrane tension $\sigma$ may vary from $10^{-8}$ to $10^{-6}$ N/m [38][39]. These values lead to $\lambda$ between 0.3 and 3 µm range, which agree with the determination of $R_{\text{eff}}$ shown in Fig. 7B as the sum of $R_P$ and $\lambda$.

Note that the degree of wrapping, local membrane deformation close to the particle, fluctuations of the equilibrium distance $d$ and weak pinning of the line at the cap edges may also play a role to explain of our experimental results [35][36].

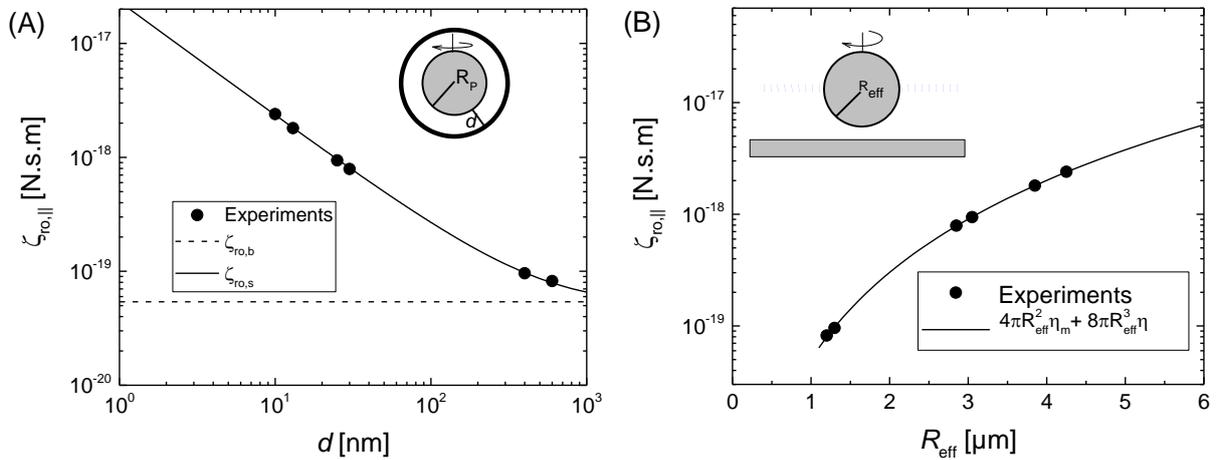

Fig. 7. Determination of $d$ (A) and $R_{\text{eff}}$ (B) by superimposing experimental data on models describing: (A) the rotational drag due to the dissipation occurring in the water gap $d$ between two concentric spheres with the outer sphere at rest; (B) rotational drag due to the dissipation of a spherical inclusion in a flat membrane with an effective radius $R_{\text{eff}}$ close to a solid wall.

Finally, we show the results of the particle translational drag parallel to the surface $\zeta_{\text{tr,||}}$, which is plotted as a function of $\zeta_{\text{ro,||}}$ in Figure 8. For two measurements of wrapped particles, $\zeta_{\text{tr,||}}$ is about 3-4 times larger than for "free" particles. For the other measurements, $\zeta_{\text{tr,||}}$ of wrapped particles is on average larger than for non-wrapped particles but it can be compared to the value of the translational drag measured for the particles closest to the wall (see Figure 3).

As for the rotational drag, also for the translational drag we are not aware of any models describing the dissipations of a particle partially wrapped by a deformed membrane and moving parallel to it, as sketched in Fig. 4A. However, it is important to notice that for the translational motion, the cap region of the particle wrapped by the membrane is not subjected to the same strong shear flows as in the rotational particle motion.

Considering the model by Evans and Sackmann for the translational drag of a disk embedded in a flat membrane close to a wall [10] and the drag of the particle close to the wall, we can write [6]:

$$\zeta_{\mathrm{tr},\parallel} \cong f_{\mathrm{w}} \, 6\pi\eta R_{\mathrm{P}} + 4\pi\eta_{\mathrm{m}} \tag{9}$$

Note that in equation 9 no specific dissipation of the membrane cap region is taken into account. In Fig. 8, we plot two calculations of equation 9 with $\eta_{\mathrm{m}} = 2 \ 10^{-9}$ Pa.s.m [14] for two different values of the particle radius (dashed line: $R_{\mathrm{P}} = 1.29$ µm, and dotted line: $R_{\mathrm{eff}} = 4$ µm) and assuming $f_{\mathrm{w}} = 1.7$, which takes into account both the effects of the vicinity to a solid wall and the particle penetration inside an inviscid interface [6]. Experimental data agree with the value of membrane surface viscosity reported in the literature [14]. Distribution of $\zeta_{\mathrm{tr},\parallel}$ experimental values can be discussed (as before in Fig. 7B) as an effective particle radius taking into account the membrane deformation close to the particle. As for the rotational friction, the largest translational drags ($\approx 1.5 \ 10^{-7}$ Pa.s.m) agree with the calculation of equation 9 with $R_{\mathrm{eff}} \approx 4$ µm. The latter $R_{\mathrm{eff}}$ value is in agreement with the sum of the particle radius and expected bendocapillary length in our system. Note that $4\pi\eta_{\mathrm{m}}$ in equation 9 describes the dissipation due to the viscosity of a flat membrane when the probe particle is in molecular contact with the membrane. Here, dissipations occur both in the membrane that shows local shape deformations close to the particle and in the water gap between the membrane and the particle [35][40].

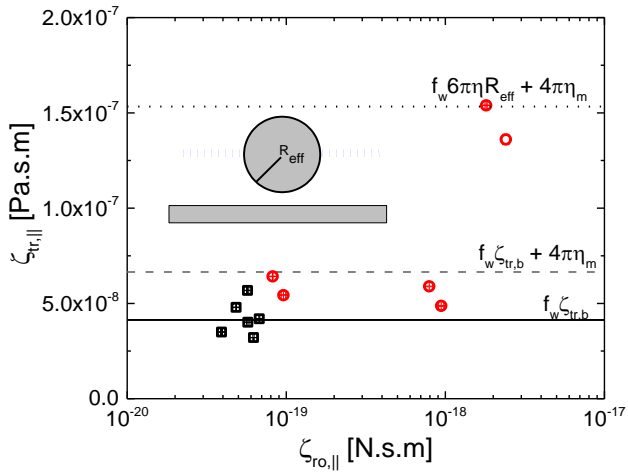

Fig. 8. Particle translational drag parallel to the bottom wall as a function of the parallel rotational drag for wrapped (circles) and "free" (squares) particles. Calculated translational drag for a sphere close to a solid wall (solid line: $\zeta_{\mathrm{tr,b}} = 6\pi\eta R_{\mathrm{P}}$, $R_{\mathrm{P}} = 1.29$ µm), and translational drags (equation 9) for two given particle radii (dashed line: $R_{\mathrm{P}} = 1.29$ µm; and dotted line: $R_{\mathrm{eff}} = 4$ µm), with $f_{\mathrm{w}} = 1.7$ and $\eta_{\mathrm{m}} = 2 \ 10^{-9}$ Pa.s.m.

# 4 Conclusions

We have shown an experimental and theoretical study to measure rotational and translational drags of single microparticles. The only experimental requirements are the use of fluorescence microscopy and Janus particles, which present clear advantages with respect to other state-of-the-art setups [41][42]. A theoretical analysis for the rotational dynamics is described, which includes a correction (Eq. 1C) that can be also applied to previously published results [36]. For an isolated particle diffusing close to a solid wall, we are able to measure the in-plane, out-of-plane rotational drags and the translational drag parallel to the wall. If the particle drags are only due to hydrodynamics, the size and the particle-wall distance can be evaluated for each particle without the need of any assumptions. Janus particles partially wrapped by lipid membranes experience rotational drags that can be 1 or 2 orders of magnitude larger than for non-wrapped particles, while the translational drag may increase up to 4 times (comparing with non-wrapped particles). In absence of strong binding interactions, the equilibrium distance between the membrane and the particle lies in the 1 to 100 nm range, which strongly affects the dissipations occurring in the particle rotational and translational dynamics. No such a distance and related geometry was taken into consideration in the existing models, which describe the probe particle in molecular contact with a flat lipid membrane. Hence, new theoretical development is demanded in this area to describe dissipations occurring in the different layers of particle-membrane systems. For sake of comparison, we evaluated the dissipations taking place in two limiting cases: (i) for an immobile membrane, we consider the dissipation of the water gap between the particle and membrane as the one occurring in the fluid between two concentric spheres, and (ii) by assuming a spherical inclusion with a relatively large effective radius to take into account the membrane deformation close to the particle. Other dissipation mechanisms may occur in the membrane cap region due to weak pinning of the contact line delimiting the wrapped and unwrapped membrane regions. For lipid bilayers, the contribution of the interleaflet friction must be also taken into consideration. Finally, we are currently performing experiments tuning the properties of the probe particle and membrane to control the equilibrium distances with the membrane and the wall and the wrapping degree.

## Authorship contribution statement

Vaibhav Sharma: Conceptualization, Methodology, Data curation, Investigation. Florent Fessler: Methodology, Data curation, Investigation. Fabrice Thalmann: Theory, Methodology, Data curation, Writing – review & editing. Carlos M Marques: Conceptualization, Methodology, Supervision, Writing – review & editing. Antonio Stocco: Conceptualization, Methodology, Data curation, Investigation, Supervision, Project administration, Resources, Writing – original draft.

## Declaration of Competing Interest

The authors declare that they have no known competing financial interests or personal relationships that could have appeared to influence the work reported in this paper.

## Acknowledgments

The authors would like to thank Marc Schmutz and the Microscopy Platform of ICS, the ITI HiFunMat (University of Strasbourg), the ANR EDEM.

## Supplementary material

The following file include the Supplementary data to this article:

Graphical Abstract

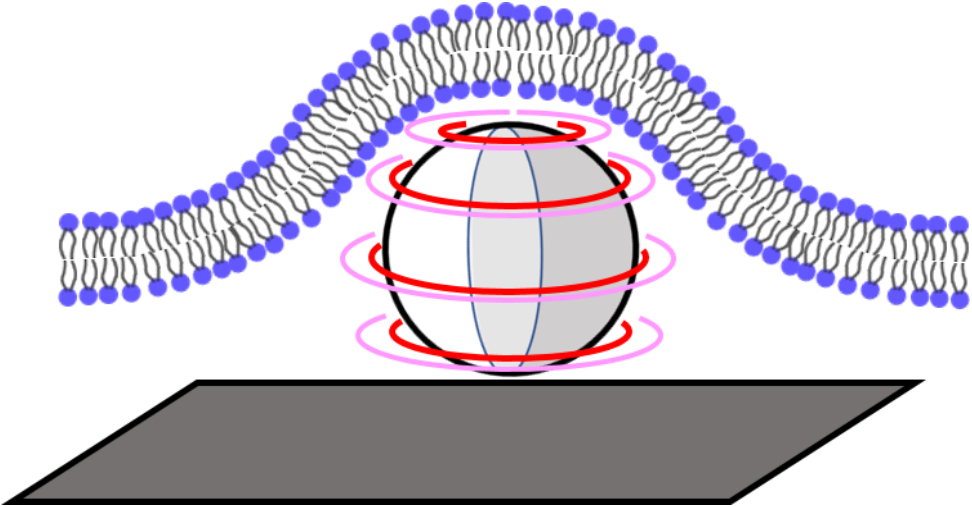